\documentclass{aastex}
\usepackage{emulateapj5}
\usepackage{apjfonts}
\usepackage{epsf}
\usepackage{graphicx}
\slugcomment{UTAP-529}
\shorttitle{Propagation of UHECRs above $10^{19}$ eV}
\shortauthors{TAKAMI ET AL.}
\begin{document}
%%%%%%%%%%%%%%%%%%%%%%%%%%%%%%%%%%%%%%%%%%%%%%%%%%%%%%%%%%%%%%%%%%%%%%
%%%%%%%%%%%%%%%%%%%%%%%%%%%%%%%%%%%%%%%%%%%%%%%%%%%%%%%%%%%%%%%%%%%%%%
  \title{Propagation OF Ultra-High Energy Cosmic Rays above $10^{19}$ eV
  in a structured extragalactic magnetic field and Galactic magnetic field}
%%%%%%%%%%%%%%%%%%%%%%%%%%%%%%%%%%%%%%%%%%%%%%%%%%%%%%%%%%%%%%%%%%%%%%
%%%%%%%%%%%%%%%%%%%%%%%%%%%%%%%%%%%%%%%%%%%%%%%%%%%%%%%%%%%%%%%%%%%%%%
%
%%%%%%%%%%%%%%%%%%%%%%%%%%%%%%%%%%%%%%%%%%%%%%%%%%%%%%%%%%%%%%%%%%%%%%
\author{Hajime Takami\altaffilmark{1}, Hiroyuki Yoshiguchi\altaffilmark{1} and Katsuhiko Sato\altaffilmark{1,2}}

\altaffiltext{1}{Department of Physics, School of Science, the University of Tokyo, 7-3-1 Hongo, Bunkyoku, Tokyo 113-0033, Japan}
\altaffiltext{2}{Research Center for the Early Universe, School of Science, the University of Tokyo, 7-3-1 Hongo, Bunkyoku, Tokyo 113-0033, Japan}

\email{takami@utap.phys.s.u-tokyo.ac.jp}
%%%%%%%%%%%%%%%%%%%%%%%%%%%%%%%%%%%%%%%%%%%%%%%%%%%%%%%%%%%%%%%%%%%%%%
%
\received{}
\accepted{}
\begin{abstract}
We present numerical simulations on propagation of Ultra-High Energy Cosmic Rays (UHECRs) above $10^{19}$ eV in a structured extragalactic magnetic field (EGMF) and simulate their arrival distributions at the earth. We use the IRAS PSCz catalogue in order to construct a model of the EGMF and source models of UHECRs, both of which reproduce the local structures observed around the Milky Way. We also consider modifications of UHECR arrival directions by the galactic magnetic field. We follow an inverse process of their propagation from the earth and record the trajectories.  This enables us to calculate only trajectories of UHECRs arriving at the earth, which saves the CPU time. From these trajectories and our source models, we construct arrival distributions of UHECRs and calculate the harmonic amplitudes and the two point correlation functions of them. We estimate number density of sources which reproduces the Akeno Ground Air Shower Array (AGASA) observation best. As a result, we find that the most appropriate number density of the sources is $\sim 5 \times 10^{-6}$ Mpc$^{-3}$. This constrains the source candidates of UHECRs. We also demonstrate skymaps of their arrival distribution with the event number expected by future experiments and examine how the EGMF affects their arrival distribution. A main result is diffusion of clustering events which are obtained from calculations in the absence of the EGMF. This tendency allows us to reproduce the observed two point correlation function better.
\end{abstract} 
\keywords{cosmic rays --- methods: numerical --- IGM: magnetic fields ---
galaxies: general --- large-scale structure of the universe}
%
%%%%%%%%%%%%%%%%%%%%%%%%%%%%%%%%%%%%%%%%%%%%%%%%%%%%%%%%%%%
%%%%%%%%%%%%%%%%%%%%%%%%%%%%%%%%%%%%%%%%%%%%%%%%%%%%%%%%%%%
\section{INTRODUCTION} \label{intro}
%%%%%%%%%%%%%%%%%%%%%%%%%%%%%%%%%%%%%%%%%%%%%%%%%%%%%%%%%%%
%%%%%%%%%%%%%%%%%%%%%%%%%%%%%%%%%%%%%%%%%%%%%%%%%%%%%%%%%%%
The nature of Ultra-High Energy Cosmic Rays (UHECRs), which are particles of energy above $10^{19}$ eV, is poorly known. This is one of the most challenging problems of modern astrophysics.

One of problems about UHECRs is what their origin is. The two scenarios of their origin are suggested, which are called bottom-up and top-down ones. On the one hand, bottom-up scenarios assume some astrophysical phenomena as their origin. UHECRs are thought to be of extragalactic origin since the gyroradii of UHECRs above $10^{19}$ eV exceed thickness of our galaxy. From this fact and the Hillas plot \citep*{hillas84}, probable candidates of UHECR origins are active galactic nuclei (AGNs), gamma-ray bursts (GRBs) and colliding galaxies. Theoretically, this scenario predicts the GZK cutoff of the energy spectrum of UHECRs \citep*{greisen66,zatsepin66} since their source candidates are located at far distances. UHE protons with energy above $\sim 4 \times 10^{19}$ eV interact with the cosmic microwave background (CMB) and lose large fraction ($\sim 20 \%$) of their energy per interaction by photopion production \citep*{berezinsky88,yoshida93}. The mean free path of UHE protons through the CMB field is $\sim$ 10 Mpc at $10^{20}$ eV. Thus the energy spectrum at the Earth should have a cutoff around $E \sim 8 \times 10^{19}$ eV. This spectral cutoff is called the GZK cutoff. But there is a observational disagreement of the energy spectra between the Akeno Giant Air Shower Array (AGASA), which does not observe the GZK cutoff \citep*{takeda98}, and the High Resolution Fly's Eye \citep*[HiRes;][]{wilkinson99}, which does it \citep*{abu02}. This discrepancy between the two experiments remains being one of open questions in astroparticle physics. On the other hand, top-down scenarios assume some processes based on new physics beyond the standard model of the particle physics (see a review \cite{bhattacharjee00}). 

\begin{figure*}
\begin{center}
\epsscale{1.3}
\plotone{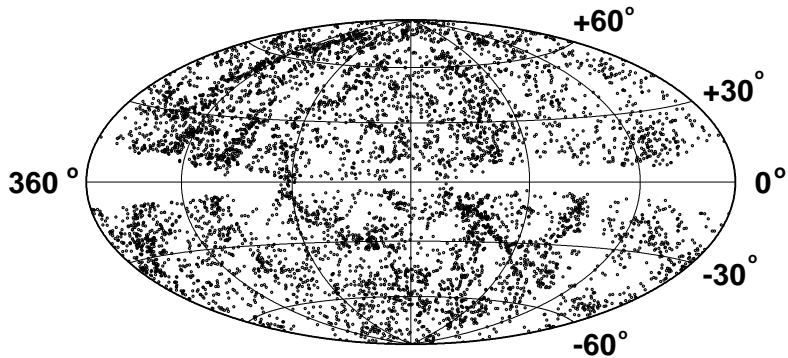} 
\caption{The IRAS PSCz galaxies within 100 Mpc on the galactic coordinate. Strong infrared luminosity on the galactic plane interrupts observation of galaxies near the galactic plane.}
\label{iras_galaxy}
\end{center}
\end{figure*}

Another problem is arrival distribution of UHECRs. The AGASA reported that there is no statistically significant large scale anisotropy in the observed arrival distribution of UHECRs above $10^{19}$ eV \citep*{takeda99}. This fact points out that sources of UHECRs are distributed isotropically, but isotropic distribution of sources cannot reproduce the small-scale anisotropy reported by the AGASA \citep*{takeda99, takeda01}. A model of UHECR origin are constrained by their ability to reproduce such observed arrival distribution of UHECRs. On the other hand, the HiRes experiment indicates that there is no statistically significant small-scale anisotropy \citep*{abbasi04b,farrar04}. However \cite*{yoshiguchi04} concluded this discrepancy between the two observations is not statistically significant at present. This problem is left for future investigation by new experiments such as the Pierre Auger Observatory. 

To obtain information on UHECR origin, we need to calculate their arrival distribution using some kinds of source models. To do so, we have to simulate propagation of UHECRs in the intergalactic space, where the Extragalactic Magnetic Field (EGMF) plays important roles since we assume that UHECRs are protons in this paper. \cite*{yoshiguchi03a} calculated propagation of UHE protons in an uniform turbulence of magnetic field with the Kolmogorov spectrum. But such magnetic field is not realistic since the EGMF is expected to reflect the large scale structure of the universe. 

In recent years, several groups started to develop physically more realistic models of the EGMF based on numerical simulations of large scale structure formation. \cite{sigl03, sigl04} used a structured EGMF model which is generated by their large scale structure simulations, taking magnetic fields into account. But their model does not reproduce the local structures actually observed around the Milky Way. This causes the ambiguity in the choice of observer position. In addition to this, the calculated arrival distribution of UHECRs does not correspond to the one expected at the earth. 

An important step on modeling the magnetic structure of the local universe is performed by \cite{dolag05}. They constrain the initial conditions for the dark matter density fluctuations to reproduce the local structures. This allows us to remove the ambiguity in the choice of observer position, and to obtain the simulated skymaps of expected UHE proton deflections in the magnetic large-scale structure around our galaxy. However, they did not calculate the arrival distribution of UHECRs, and thus could not obtain the information on source distribution which reproduces the AGASA observation. And also, the effects of the Galactic Magnetic Field (GMF) are not considered in both \cite{sigl03,sigl04} and \cite{dolag05}. But recently it has also been shown that the GMF affects the arrival distribution of UHECRs \citep*{stanev02, yoshiguchi03d}. Thus we cannot neglect modifications of arrival directions of UHECRs by the GMF when we simulate their arrival distribution.

In this work, we calculate propagation of UHECRs taking both the EGMF and the GMF into account, and simulate the arrival distribution of UHECRs. We constrain source number density of UHECRs by comparison of the results with the AGASA observation. We generate the magnetic structure of the local universe by our original method (section 3.1) from the IRAS PSCz catalogue of galaxies \citep*{saunders00}. We also construct our source models of UHECRs from this catalogue. As our GMF model, we adopt a bisymmetric spiral field (BSS) model \citep*{stanev02} just like \cite*{yoshiguchi03d}.

In order to simulate the arrival distribution of UHECRs, we apply a method developed in previous works. \citep*{fluckiger91, bieber92, stanev96, tanco99, yoshiguchi03d}. We numerically calculate an inverse process of propagation of UHE protons, which reach the earth, and record their trajectories in our Galaxy and the intergalactic space. In other words, we inject UHECRs from the earth isotropically whose charges are taken as -1. We then select some of them according to a given source distribution. (Detailed explanation is given in the section \ref{calc_arrival}) The expected arrival distribution can be obtained by mapping the velocity directions of the selected trajectories at the earth. The validity of this method is supported by the Liouville's theorem. This method enables us to save the CPU time effectively since we calculate only trajectories of UHE protons which reach the earth. A method for this process is explained in section \ref{method}.

The outline of this paper is as follows. In section \ref{galaxies}, we explain the IRAS PSCz Catalogue and construct our sample of galaxies. In section \ref{mf}, we introduce our model of the EGMF and the GMF. We explain our numerical methods for calculating arrival distribution of UHECRs and statistical quantities in section \ref{method}. Then in section \ref{result}, we estimate the most appropriate number density of source of UHECRs and compare the statistical quantities calculated from this source model with the EGMF to those calculated without the EGMF. We also demonstrate skymaps of the arrival distribution of UHECRs. In section \ref{summary}, we summarize the main results.

%%%%%%%%%%%%%%%%%%%%%%%%%%%%%%%%%%%%%%%%%%%%%
%%%%%%%%%%%%%%%%%%%%%%%%%%%%%%%%%%%%%%%%%%%%%
\section{A SAMPLE OF GALAXIES} \label{galaxies}
%%%%%%%%%%%%%%%%%%%%%%%%%%%%%%%%%%%%%%%%%%%%%
%%%%%%%%%%%%%%%%%%%%%%%%%%%%%%%%%%%%%%%%%%%%%
In order to calculate propagation of UHECRs considering the local structures actually observed around the Milky Way, we use the IRAS PSCz Catalogue \citep*{saunders00} for construction of our EGMF model and UHECR source models.

We used the ORS sample of galaxies to construct our UHECR source models in our previous work \citep*{yoshiguchi03a}. The ORS sample has better completeness on nearby galaxies than the IRAS PSCz catalogue. Thus we used the ORS sample to construct our source model since we were interested in the local sources in order to research nature of UHECRs above $4 \times 10^{19}$ eV. As a result, we obtained that UHECR source number density is $\sim 10^{-6} \mathrm{Mpc}^{-3}$. Then we investigated that this source model also explained arrival distribution of UHECRs above $10^{19}$ eV though the EGMF was neglected \citep*{yoshiguchi03d}. In this work, we use galaxy sample to construct not only a source model of UHECRs but also a model of the EGMF, which reflect the large scale structure of the universe. This requests large sky coverage of the galaxy sample. Thus we adopt not the ORS galaxy catalogue but the IRAS PSCz Catalogue.

The IRAS PSCz catalogue consists of 14677 galaxies with redshift and infrared fluxes $>$ 0.6 Jy, and covers about 84\% of the sky. We assume de Sitter universe with $\Omega_{m} = 0.3, \Omega_{\lambda} = 0.7, H_0 = 71$ $\mathrm{km}~\mathrm{s}^{-1}~\mathrm{Mpc}^{-1}$ in order to calculate distance of each galaxy. We show distribution of the IRAS PSCz galaxies in Figure \ref{iras_galaxy}.

However, there are two problems when we use the IRAS PSCz Catalogue. One is that it is impossible to observe dark galaxies at far-off distance. (the selection effect, see figure \ref{ld_relation}). The other is that this catalogue has the zone of avoidance (the mask), where the IRAS PSCz Survey did not observe galaxies. Our previous works \citep*{yoshiguchi03a, yoshiguchi03d}, which uses the ORS sample \citep*{santiago95}, also have these problems. Using luminosity function of the IRAS galaxies, we correct these absence of galaxies in the same manner with our previous studies. 

We use the luminosity function of the IRAS PSCz galaxies \citep*{takeuchi03}, 

  \begin{equation}
    \Phi(L) = \phi_* \left( \frac{L}{L_*} \right)^{1-\alpha} 
    \exp \left\{ -\frac{1}{2\sigma^2} \left[ \log \left( 1 + 
      \frac{L}{L_*} \right) \right]^2 \right\}.
  \end{equation} 
Here $L_* = (4.34 \pm 0.86 )\times 10^8 h^{-2} [L_{\odot}], \alpha = 1.23 \pm 0.04, \sigma = 0.724 \pm 0.010, \phi_* = (2.34 \pm 0.30) \times 10^{-2} h^3 \mathrm{Mpc}^3$. Using this luminosity function, we define the selection function as,

  \begin{equation}
    \phi(r) = \frac{\int_{L_{\mathrm{min}(r)}}^{\infty} dL \Phi(L)}
	{\int_0^{\infty} dL \Phi(L)},
  \end{equation}
where $L_{\mathrm{min}(r)}$ is minimum luminosity of galaxies which are observable at a distance $r$. Therefore, $\phi(r)$ represents fraction of all galaxies that are observable at each distance.

First of all, we correct the selection effect. For each of the IRAS galaxies, we add galaxies that are not included in the IRAS sample. The number of added galaxies can be obtained by using the selection function. The positions of added galaxies are determined according to the Gaussian distribution whose mean is the position of the original IRAS galaxy and whose root mean square is $l(r)$. Here we define $l(r)$ as a mean distance between the original IRAS galaxies at distance $r$, thus 

\begin{equation}
 \frac{4 \pi}{3} {l(r)}^3 n(r) = 1, 
\end{equation}
where $n(r)$ is number density of the original IRAS galaxies. Then,
\begin{equation}
 l(r) = \left( \frac{3}{4 \pi} \right)^{1/3} n(r)^{-1/3}.
\end{equation}

The luminosities of added galaxies are randomly assigned so that their distribution of luminosity is consistent with the luminosity function. This method can complement the IRAS galaxies without spoiling the observed structure of galaxy distribution.

\vspace{0.5cm}
\centerline{{\vbox{\epsfxsize=8.0cm\epsfbox{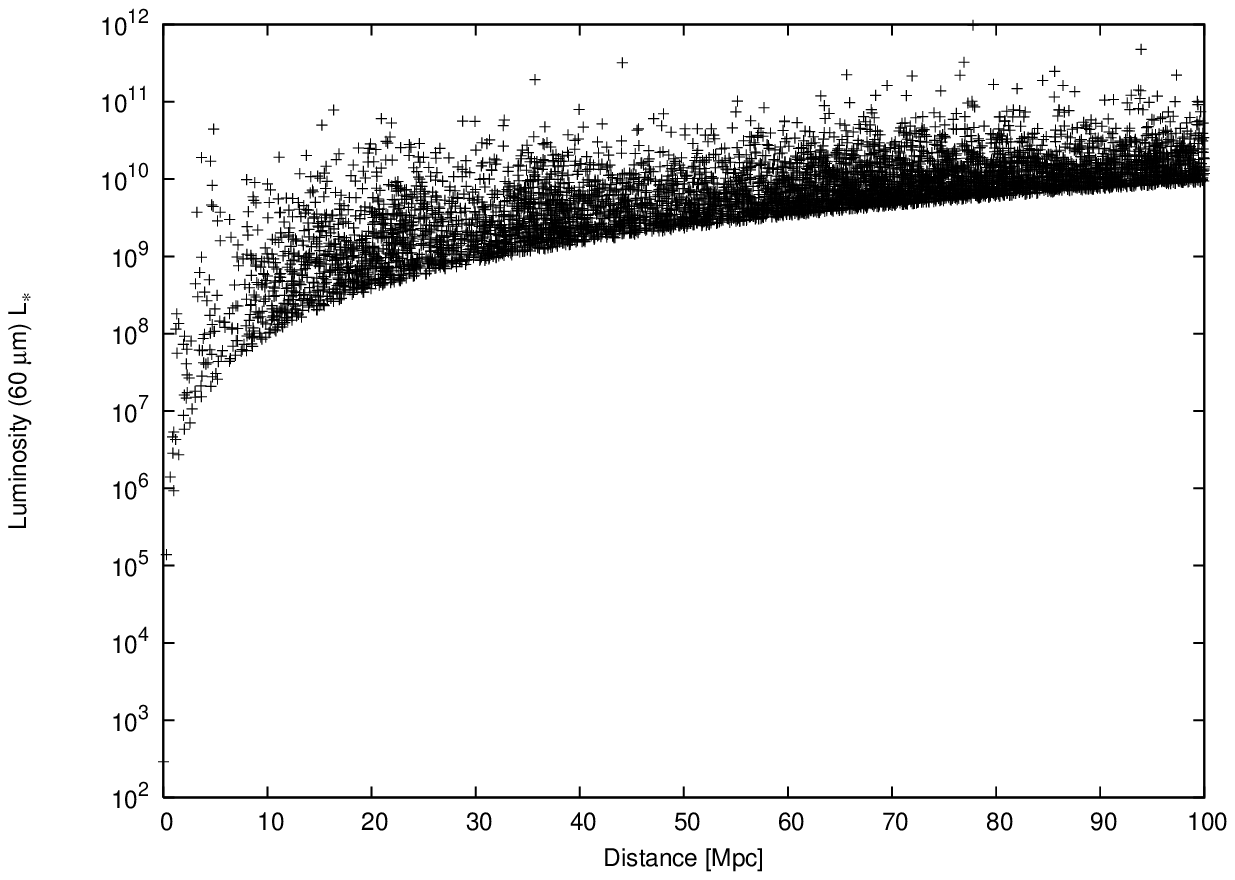}}}}
\figcaption{Infrared luminosity (60$\mu$m) of the IRAS PSCz galaxies
plotted as a function of distance from the Earth. Unit of luminosity is
$\mathrm{L}_{\odot}$.\label{ld_relation}}
\vspace{0.5cm}

\begin{figure*}
\begin{center}
\epsscale{1.3}
\plotone{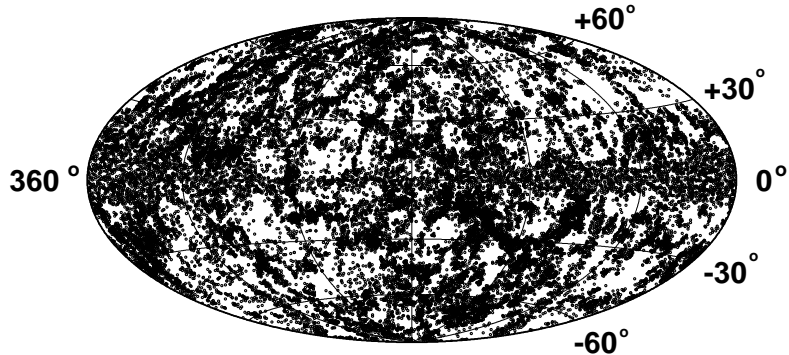} 
\caption{Distribution of galaxies after our correction for the selection effect and existance of the mask. This distribution do not spoil the observed structure of galaxy distribution.}
\label{corrected_galaxy}
\end{center}
\end{figure*}

Next, we add galaxies in the mask. We assume that galaxy distribution in this region is homogeneous and number density of galaxies is $n(r) \phi(r)^{-1}$. These luminosities are random but their distribution is consistent with the luminosity function. Our galaxy sample after these corrections is shown in Figure \ref{corrected_galaxy}.

In this work, we use only the IRAS galaxies within $100$ Mpc. We assume that source distribution at $r > 100$ Mpc is isotropic and uniform, and that their number density is equal to that within $100$ Mpc. We neglect cosmological evolution of number density of galaxies. Thus our sample of galaxies reflects the observed local structure within 100 Mpc. We use this sample of galaxies to construct a model of EGMF and source model of UHECRs.

%%%%%%%%%%%%%%%%%%%%%%%%%%%%%%%%%%%%%%%%%%%%%
%%%%%%%%%%%%%%%%%%%%%%%%%%%%%%%%%%%%%%%%%%%%%
\section{A MODEL OF MAGNETIC FIELD} \label{mf}
%%%%%%%%%%%%%%%%%%%%%%%%%%%%%%%%%%%%%%%%%%%%%
%%%%%%%%%%%%%%%%%%%%%%%%%%%%%%%%%%%%%%%%%%%%%

%%%%%%%%%%%%%%%%%%%%%%%%%%%%%%%%%%%%%%%%%%%%%
\subsection{Extragalactic Magnetic Field} \label{egmf}
%%%%%%%%%%%%%%%%%%%%%%%%%%%%%%%%%%%%%%%%%%%%%

The EGMF are little known theoretically and observationally. Theoretically, several large scale structure simulations with magnetic field have been performed \citep*{dolag05, sigl04}. Roughly speaking, their results are that the strength of magnetic field traces baryon density. A model of \cite*{sigl04} do not reflect the local structures actually observed around the Milky Way, while one of \cite*{dolag05} reflects these local structures. To compare model predictions of UHECR arrival distribution with the observed one, it is important to generate the magnetic structure around our galaxy. Therefore, we present a model of the EGMF reflecting the local structures of the universe well. 

The EGMF mainly exists in clusters of galaxy or around galaxies. From this standpoint, we present a realistic model of the EGMF. We assume that magnetic field results from the amplification of weak seed fields. In a simulation of evolution of cluster of galaxy \citep*{dolag02}, the average magnetic field strength in the cluster is amplified as expected from compression alone ($|B| \propto \rho^{2/3}$, where $\rho$ is density of matter).
We adopt this conclusion 

\begin{equation}
|B| \propto \rho^{2/3} \propto {\rho_L}^{2/3},  \label{221}
\end{equation} 
where $\rho_L$ is luminosity density explained in the next paragraph. In equation \ref{221}, we assume that the luminosity density on each position is proportional to density of the gas. With these assumption, we construct a model of the EGMF as follows.

First, we cover the universe with cubes of side $l_c$, which is correlation length of the EGMF. We adopt $l_c = 1 \mathrm{Mpc}$ from our previous work \citep*{yoshiguchi03a}. Second, we sum luminosities of galaxies in our sample which exist in each cube. We call this summed value luminosity density. Finally, It is assumed that the magnetic field in each cube is represented as the Gaussian random field with zero mean and has a power-law spectrum

\begin{eqnarray}
  \left< B(\overrightarrow{k})B^{*}(\overrightarrow{k}) \right> 
  &\propto& k^{n_H} \quad \mathrm{for} \quad 2\pi/l_c \leq k 
  \leq 2\pi/l_{\mathrm{cut}}, \nonumber \\
  \left< B(\overrightarrow{k})B^{*}(\overrightarrow{k}) \right> &=& 0
  \quad \mathrm{otherwise}, 
\end{eqnarray}
where $l_{\mathrm{cut}}$ is a numerical cutoff scale. Physically, one expects $l_{\mathrm{cut}} \ll l_{\mathrm{c}}$, but we set $l_{\mathrm{cut}} = 1/8 \times l_{\mathrm{c}}$ in order to save the CPU time. We use $n_{\mathrm{H}} = -11/3$, corresponding to the Kolmogorov spectrum since the Faraday rotation map reveals that the clusters' magnetic fields are turbulent with the Kolmogorov spectrum over at least one order of magnitude of the wavenumber\citep*{vogt04}.

Next we consider a normalization of the EGMF. Most observations suggest that clusters of galaxy have magnetic field whose strength is from 0.1 $\mu$G to a few $\mu$G (see a review \cite*{vallee04}). On the one hand, the Faraday rotation measurements of polarized radio sources placed within cluster of galaxies provide some evidence for the presence of stronger intracluster magnetic field (ICMF), in the range of a few $\mu$G \citep*{taylor01, vogt03}. On the other hand, observations of hard X-ray emission from cluster of galaxy implies that an average ICMF strength within the emitting volume is 0.2-0.4 $\mu$G \citep*{fusco99, rephaeli99}. 

\vspace{0.5cm}
\centerline{{\vbox{\epsfxsize=8.0cm\epsfbox{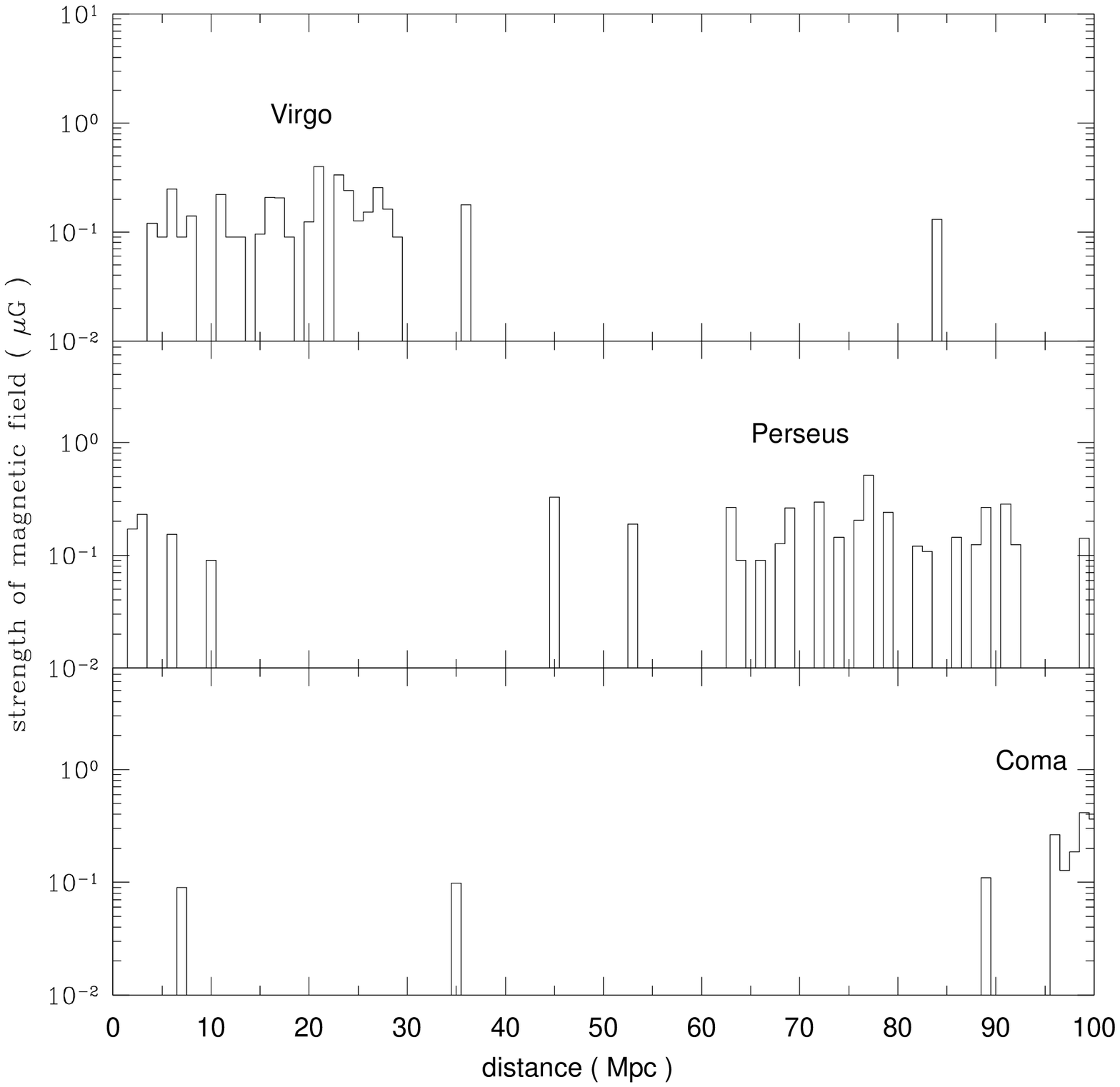}}}}
\figcaption{Magnetic field strength along three fiducial lines through the Virgo cluster, the Perseus cluster and the Coma cluster within 100 Mpc.\label{sliced_mf}}
\vspace{0.5cm}

\begin{figure*}
\begin{center}
\epsscale{1.5}
\plotone{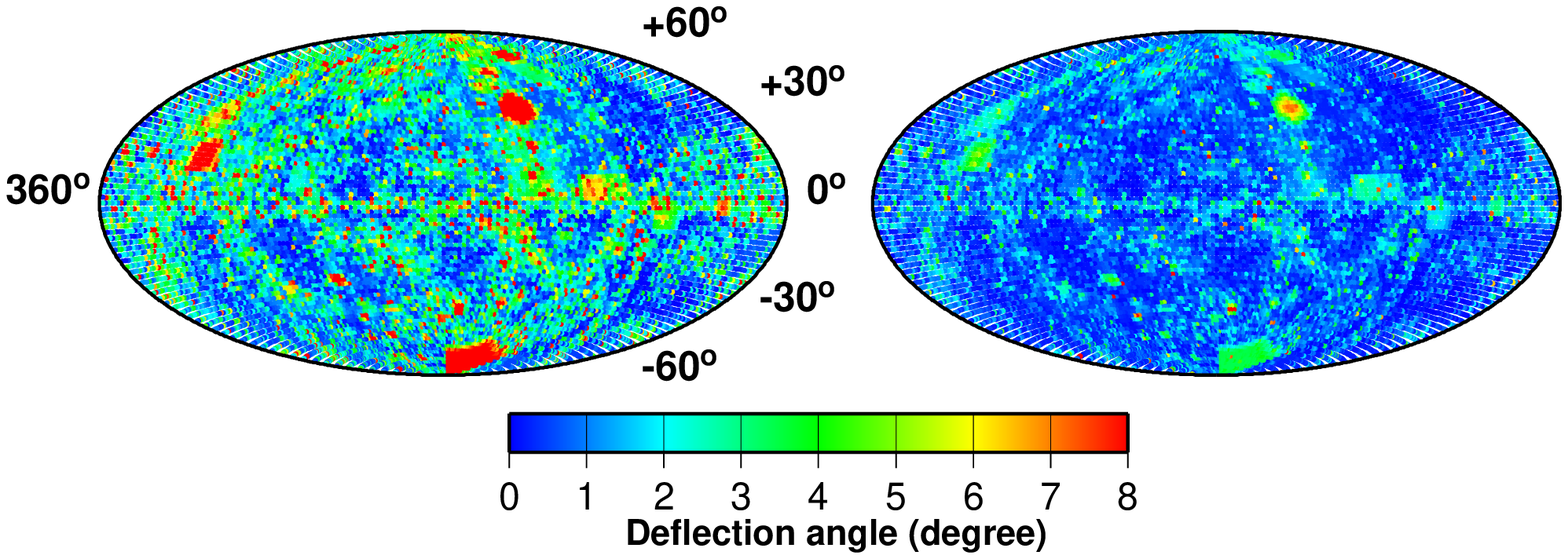} 
\caption{The deflection angles of UHE protons with arrival energy $E = 4 \times 10^{19} \mathrm{eV}$ ( left panel ) and $E = 10^{20} \mathrm{eV}$ ( right panel ) through propagation in our EGMF from the distance of 100 Mpc.}
\label{projection_mf}
\end{center}
\end{figure*}

In this work, we now set a normalization of its strength $\sim 0.4 \mu$G in a cube where is the center of the Virgo cluster. In order to compare our model with \cite*{dolag05}, we show magnetic field strength within $100$ Mpc along three fiducial lines through the Virgo cluster, the Perseus cluster and the Coma cluster in figure \ref{sliced_mf}. In figure \ref{projection_mf}, we also show deflection maps when protons propagate through our EGMF from the distance of 100 Mpc. The deflection angle by uniform turbulent magnetic field is given by 
\begin{equation}
\theta \simeq 0.3^{\circ} \left( \frac{E}{10^{20} \mathrm{eV}} \right)^{-1} \left( \frac{r}{100 \mathrm{Mpc}} \right)^{1/2} \left( \frac{l_c}{1 \mathrm{Mpc}} \right)^{1/2} \left( \frac{B}{10^{-4} \mathrm{\mu G}} \right). 
\end{equation} 
Here $r$ is distance of propagation. In the way similar to \cite{dolag05}, we assume that proton trajectory makes a random walk through each cell since our EGMF in each cube of side $l_c$ is the turbulence. These maps can be compared to figure 13 and 14 in \cite{dolag05}. 

We use our sample of galaxies to construct a model of the EGMF only within $100$ Mpc. At a distance above $100$ Mpc, we treat the EGMF as an uniform turbulence of magnetic field with the same spectrum and $|B|$ = $1$ nG since the IRAS PSCz catalogue is poorly covered at $r > 100$ Mpc.

%%%%%%%%%%%%%%%%%%%%%%%%%%%%%%%%%%%%%%%%%%%%%
\subsection{Galactic Magnetic Field} \label{gmf}
%%%%%%%%%%%%%%%%%%%%%%%%%%%%%%%%%%%%%%%%%%%%%

In this study, we adopt the GMF model used by \cite{stanev02}, which is composed of the spiral and the dipole field. We briefly explain this GMF model below. 

Faraday rotation measurements indicate that the GMF in the disk of the Galaxy has a spiral structure with field reversals at the optical Galactic arms \citep{beck01}. We use a bisymmetric spiral field (BSS) model, which is favored from recent work \citep{han99,han01}. The Solar System is located at a distance
$r_{\vert\vert}=R_\oplus=8.5$ kpc from the center of the Galaxy in the Galactic plane. The local regular magnetic field in the vicinity of the Solar System
is assumed to be $B_{\rm Solar} \sim 1.5~\mu{\rm G}$ in the direction $l=90^{\rm o}+p$ where the pitch angle is $p=-10^{\rm o}$ \citep{han94}.

In the polar coordinates $(r_{\vert\vert},\phi)$, the strength of the spiral field in the Galactic plane is given by

\begin{equation}
B(r_{\vert\vert},\phi)=
B_0~\left({R_\oplus \over r_{\vert\vert}}\right)~
\cos\left(\phi - \beta \ln {r_{\vert\vert}\over r_0} \right),
\end{equation}  
where $B_0=4.4~\mu$G, $r_0= 10.55$ kpc and $\beta=1/\tan p=-5.67$. The field decreases with Galactocentric distance as $1/r_{\vert\vert}$ and it becomes zero for $r_{\vert\vert}>20$ kpc. In the region around the Galactic center ($r_{\vert\vert} < 4$ kpc) the field is highly uncertain, and thus assumed to be constant and equal to its value at $r_{\vert\vert}=4$ kpc.

The spiral field strengths above and below the Galactic plane are taken to decrease exponentially with two scale heights \citep{stanev96}, 

\begin{equation}
\vert B(r_{\vert\vert},\phi,z)\vert = 
\vert B(r_{\vert\vert},\phi)\vert 
\left\{
\begin{array}{lcl}
\exp(-|z|) :  & \vert z \vert \leq 0.5~ {\rm kpc} \\ 
\exp( \frac{-3}{8})~\exp(\frac{-|z|}{4}) :  & \vert z \vert > 0.5~ {\rm kpc} 
\end{array}
\right.
\label{eq:b-height}
\end{equation} 
where the factor $\exp(-3/8)$ makes the field continuous on $z$. The BSS spiral field we use is of even parity, that is, the field direction is preserved at disk crossing. 

Observations show that the field in the Galactic halo is much weaker than that in the disk. In this work we assume that the regular field corresponds to a A0 dipole field as suggested in \citep{han02}. In spherical coordinates $(r,\theta,\varphi)$, the $(x,y,z)$ components of the halo field are given by:

\begin{eqnarray}
B_x=-3~\mu_{\rm G}~{\sin\theta \cos\theta \cos\varphi}/
r^3 \nonumber \\ 
B_y=-3~\mu_{\rm G}~{\sin\theta \cos\theta \sin\varphi}/
r^3 \\
B_z=\mu_{\rm G}~{(1-3\cos^2\theta)}/r^3 ~~~~~~~ \nonumber
\label{eq:bdipole}
\end{eqnarray}
where $\mu_{\rm G}\sim 184.2~{\rm \mu G~kpc^3}$ is the magnetic moment of the Galactic dipole. The dipole field is very strong in the central region of the Galaxy, but is only 0.3 $\mu$G in the vicinity of the Solar system, directed toward the North Galactic Pole.

There is a significant turbulent component, $B_{\rm random}$, of the Galactic magnetic field. Its field strength is difficult to measure but results found in
literature are in the range of $B_{\rm random} = 0.5 \dots 2 B_{\rm reg}$ \citep{beck01}. However, we neglect the random field through the paper. Possible dependence of the results on the random field is discussed in the section~\ref{result}.

%%%%%%%%%%%%%%%%%%%%%%%%%%%%%%%%%%%%%%%%%%%%%
%%%%%%%%%%%%%%%%%%%%%%%%%%%%%%%%%%%%%%%%%%%%%
\section{NUMERICAL METHOD} \label{method}
%%%%%%%%%%%%%%%%%%%%%%%%%%%%%%%%%%%%%%%%%%%%%
%%%%%%%%%%%%%%%%%%%%%%%%%%%%%%%%%%%%%%%%%%%%%

%%%%%%%%%%%%%%%%%%%%%%%%%%%%%%%%%%%%%%%%%%%%%%%
\subsection{Method of Calculation for Propagation of UHECRs} \label{sim_method}
%%%%%%%%%%%%%%%%%%%%%%%%%%%%%%%%%%%%%%%%%%%%%%%
We numerically calculate an inverse process of propagation of UHE protons arriving at the earth in the intergalactic space. This method explained below is an expansion of many previous works on their propagation in the Galactic space \citep*{fluckiger91, bieber92, stanev96, tanco99, yoshiguchi03d}. 

We already performed numerical simulations for UHECR propagation in the GMF in \cite*{yoshiguchi03d}. In the paper, we injected UHECRs from the earth isotropically, and recorded these trajectories until they reached a sphere of radius $40$ kpc centered at the Galactic center. The charge of UHECRs was taken as $-1$ because we followed propagation of UHE protons backward. These UHECRs were injected with spectral index of $-2.7$, which was similar to the observed one. Note that this is not the energy spectrum injected at extragalactic sources. 

In this study, we expand these trajectories to the extragalactic space. In other words, there are our initial positions of UHECRs on a sphere of radius $40$ kpc centered at the Galactic center, which are the result of our previous work. The trajectories are followed until their distance from the Galaxy reaches 1 Gpc or their time for propagation reaches the age of the universe or their energies reach $10^{25}$ eV. Of course, we set the charge of UHECRs to be $-1$. 

In the extragalactic space, we have to consider not only the deflections due to the EGMF but also the energy loss processes \citep*{berezinsky88, yoshida93}. UHE protons below $4 \times 10^{19}$ eV lose their energies mainly due to adiabatic energy losses and pair production in collision with the cosmic microwave background (CMB). At the higher energies the photopion production with the CMB becomes essential (Detail explanation is given below). Though we assume that UHECRs are protons in this work, we should also add the photo-disintegration if we assume UHECRs to be nuclei. We treat all these energy loss processes as continuous processes. Note that energies of UHECRs increase during propagation because we follow their inverse processes. 

The adiabatic energy loss is the effect of the expanding universe. This energy loss is written as

\begin{equation}
  \frac{d E}{dt} = -\frac{\dot{a}}{a} E 
  = - H_0 \left[ \Omega_m (1+z)^3 + \Omega_{\Lambda} \right] E .
\end{equation} 
As mentioned in the section \ref{galaxies}, the cosmological parameters used in this calculation are $\Omega_m = 0.3$, $\Omega_{\lambda} = 0.7$, and $H_0 = 71~\mathrm{km}~\mathrm{s}^{-1}~\mathrm{Mpc}^{-1}$.

The pair production due to collisions with the CMB can be treated as a continuous process which has small inelasticity ($\sim 10^{-3}$). We adopt the analytical fit functions given by \cite{chodorowski92} to calculate the energy loss rate on isotropic photons.

UHE protons above $\sim 4 \times 10^{19}$ eV lose a large fraction of their energy ($\sim 20 \%$ at every reaction) in the photopion production with the CMB. We treated this process as a stochastic process in previous work. But in this study, we cannot treat this as a stochastic process because we calculate the inverse process. \cite*{berezinsky02} and \cite*{aloisio04} showed that the energy spectrum which is calculated with a continuous process of the photopion production is consistent with one calculated with a stochastic process if a distance between the earth and sources of UHECR is more than $30$ Mpc. Thus, we can adopt a continuous energy loss process since our source model (explained below) almost satisfy that condition about source distance. We use the energy loss length which is calculated by simulating the photopion production with the event generater SOPHIA \citep*{sophia00}.

%-----------------------------------------------------------------
\subsection{Source Distribution}
\label{method_source}
%-----------------------------------------------------------------

We construct source models of UHECRs from our sample of galaxies explained in the section \ref{mf}. The number density of UHECR sources is taken as our model parameter. For a given number density, we randomly select galaxies from our sample with probability proportional to absolute luminosity of each galaxy. We then estimate the source number density which reproduces the observed arrival distribution of UHECRs, by calculating the harmonic amplitude and the two point correlation function of arrival distribution of UHECRs as a function of the source number density.

%-----------------------------------------------------------------
\subsection{Calculation of the UHECR Arrival Distribution}
\label{calc_arrival}
%-----------------------------------------------------------------

In this subsection, we explain the method of construction of UHECR arrival distribution at the earth. 

We calculate 500,000 trajectories of UHE protons in the EGMF, using our method explained in Section \ref{sim_method},  and record them. With our source models, we calculate a factor for each trajectory, which represents a relative probability that $j$ th proton reaches the Earth, 
\begin{equation}
P_{\rm selec} (E, j) \propto \sum _{i} \frac{L_{i,j}}{( 1 + z_{ij} ){d_{i,j}}^2} \, 
\frac{dN/dE_g(d_{i,j},E_i)}{E^{-2.7}} \frac{d E_g}{dE}.
\label{def_pselec} 
\end{equation}
Here, $i$ labels sources on each trajectory. $z_{i,j}$, $d_{i,j}$ and $L_{i,j}$ is a redshift, a distance and luminosity of each source which is passed by $j$ th proton respectively. $dN/dE(d_{i,j},E_i)$ is the energy spectrum of protons at a source of distance $d_{i,j}$. $E_i$ is energy of proton at $i$ th source. $E_g = E_g (E, d)$ is the energy of cosmic ray at a source, which has the energy $E$ at the earth. $d E_g / dE$ represents variation of shape of the energy spectrum through propagation. 

$d E_g / dE$ can be calculated in the case of rectilinear propagation \citep*{berezinsky88}. But calculation of this is difficult in this study since protons injected from the sources which are located at the same distance have different path length due to the EGMF. That is, the only $E_g$ cannot be decided when $E$ and $d$ are given. We calculate this factor using our 500,000 trajectories of protons. 

Figure \ref{spectrum_shift} shows a variation of shape of a monoenegetic spectrum ($E=10^{19.6}$ eV) at the earth for an example. The solid histogram is the spectrum at the earth. The dashed histogram and the dotted histogram are the spectra of UHECRs injected from the earth at 300 Mpc from us and 500 Mpc respectively. It is difficult to determine $d E_g$ from the figure since the spectra at far distances from the earth have large variances due to difference of their path lengthes. In this case, we calculate $d E_g / dE$ as 
\begin{equation}
\frac{d E_g}{dE} (E, d) = \frac{dN/dE (E)}{dN/d{E_g} ({E_g}^*(E, d), d)}, 
\end{equation}
where $dN/dE (E)$ is the spectrum at the earth ($\propto E^{-2.7}$), $dN/d{E_g} (E_g(E, d), d)$ is a spectrum of UHECRs injected from the earth at a distance $d$ and ${E_g}^*(E, d)$ is averaged $E_g$ when $E$ and $d$ are given. 

We randomly select several trajectories according to these relative probabilities, so that the number of the selected trajectories is equal to the observed event number. The mapping of the velocity directions of each UHECR at the earth becomes the arrival distribution of UHECRs. The validity of this method is supported by the Liouville's theorem. 

If we have to select the same trajectory more than once in order to adjust the number of the selected trajectories, we generate a new event whose arrival angle is calculated by adding a normally distributed deviate with zero mean and variance equal to the experimental resolution $2.8^{\circ}$ $(1.8^{\circ})$ for $E>10^{19}$ eV $(4 \times 10^{19} {\rm eV})$ to the original arrival angle. 

We assume that UHECRs are protons injected with a power law spectrum in the range of $10^{19} - 10^{22}$ eV. We set this power law index 2.6 in order to fit the calculated energy spectrum to the one observed by the AGASA \citep{marco03}. In other words, $dN/dE(d_{i,j}, E) \propto E^{-2.6}$ where $E$ is the energy of UHECR at the source.

\vspace{0.5cm}
\centerline{{\vbox{\epsfxsize=5.5cm\rotatebox{-90}{\epsfbox{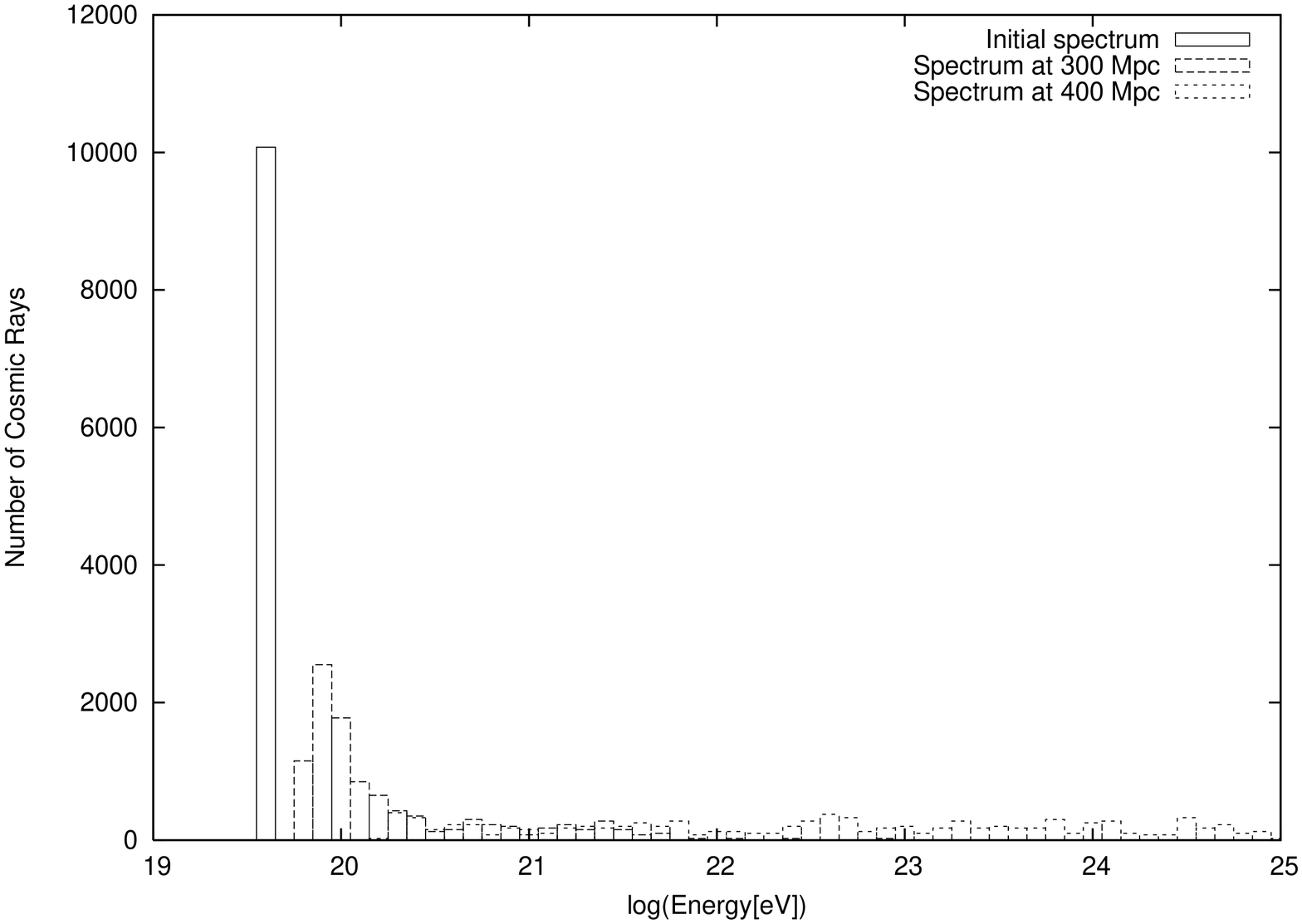}}}}}
%\rotatebox{-90}{\plotone{f6.eps}} 
\figcaption{A variation of shape of a monoenegetic spectrum ($E=10^{19.6}$ eV) at the earth through UHECR propagation. The solid histogram is the spectrum at the earth. The dashed histogram and the dotted histogram are the spectrum of UHECRs injected from the earth at 300 Mpc and at 500 Mpc from us respectively. The spectra at far distances from the earth have large variances due to difference of their path lengthes.\label{spectrum_shift}}
\vspace{0.5cm}

%-------------------------------------------------------------------------
\subsection{Statistical Methods}\label{statistics}
%-------------------------------------------------------------------------

In this subsection, we explain the two statistical quantities, the harmonic analysis for large scale anisotropy \citep*{hayashida99} and the two point correlation function for small scale anisotropy.

The harmonic analysis to the right ascension distribution of events is the conventional method to search for large scale anisotropy of cosmic ray arrival distribution. For a ground-based detector like the AGASA, the almost uniform observation in right ascension is expected. The $m$-th harmonic amplitude $r$ is determined by fitting the distribution of cosmic rays to a sine wave with period $2 \pi /m$. For a sample of $n$ measurements of phase, $\phi_1$, $\phi_2$, $\cdot \cdot \cdot$, $\phi_n$ (0 $\le \phi_i \le 2 \pi$), it is expressed as

\begin{equation}
r = (a^2 + b^2)^{1/2}
\label{eqn121}
\end{equation}
where, $a = \frac{2}{n} \Sigma_{i = 1}^{n} \cos m \phi_i  $, $b = \frac{2}{n} \Sigma_{i = 1}^{n} \sin m \phi_i  $. We calculate the harmonic amplitude for $m=1,2$ from a set of events generated by the method explained in the section~\ref{calc_arrival}.

If events with total number $n$ are uniformly distributed in right ascension, the chance probability of observing the amplitude $\ge r$ is given by,

\begin{equation}
P = \exp (-k),
\label{eqn13}
\end{equation}
where

\begin{equation}
k = n r^2/4.
\label{eqn14}
\end{equation} 
The current AGASA 775 events above $10^{19}$ eV is consistent with isotropic source distribution within 90 $\%$ confidence level \citep*{takeda01}. We therefore compare the harmonic amplitude for $P = 0.1$ with the model prediction.

The two point correlation function $N(\theta)$ contains information on the small scale anisotropy. We start from a set of events generated from our simulation. For each event, we divide the sphere into concentric bins of angular size $\Delta \theta$, and count the number of events falling
into each bin. We then divide it by the solid angle of the corresponding bin, that is,

\begin{eqnarray}
N ( \theta ) = \frac{1}{2 \pi | \cos \theta  - \cos (\theta + \Delta \theta)
|} \sum_{ \theta
\le  \phi \le \theta + \Delta \theta }  1 \;\;\; [ \rm  sr ^{-1} ],
\label{eqn100}
\end{eqnarray}
where $\phi$ denotes the separation angle of the two events. $\Delta \theta$ is taken to be $1^{\circ}$ in this analysis. The AGASA data shows correlation at small angle $(\sim 3^{\circ})$ with 2.3 (4.6) $\sigma$ significance of deviation from an isotropic distribution for $E>10^{19}$ eV $(E>4 \times 10^{19} {\rm eV})$ \citep*{takeda01}.

%%%%%%%%%%%%%%%%%%%%%%%%%%%%%%%%%%%%%%%%%
%%%%%%%%%%%%%%%%%%%%%%%%%%%%%%%%%%%%%%%%%
\section{RESULTS} \label{result}
%%%%%%%%%%%%%%%%%%%%%%%%%%%%%%%%%%%%%%%%%
%%%%%%%%%%%%%%%%%%%%%%%%%%%%%%%%%%%%%%%%%

In this section, we present results of our simulations. In section~\ref{constraint}, we constrain number density of UHECR sources from the observational results of the AGASA. Using our source model with this number density, we see how the EGMF affects the arrival distribution of UHECRs in section~\ref{arrival} and section~\ref{res_statistics}.

%%%%%%%%%%%%%%%%%%%%%%%%%%%%%%%%%%%%%%%%%
\subsection{A constraint on source model of UHECRs} \label{constraint}
%%%%%%%%%%%%%%%%%%%%%%%%%%%%%%%%%%%%%%%%%

In this subsection, we constrain source number density of UHECRs from the arrival distribution obtained by the AGASA. 

\begin{figure*}
\begin{center}
\epsscale{1.2}
\plotone{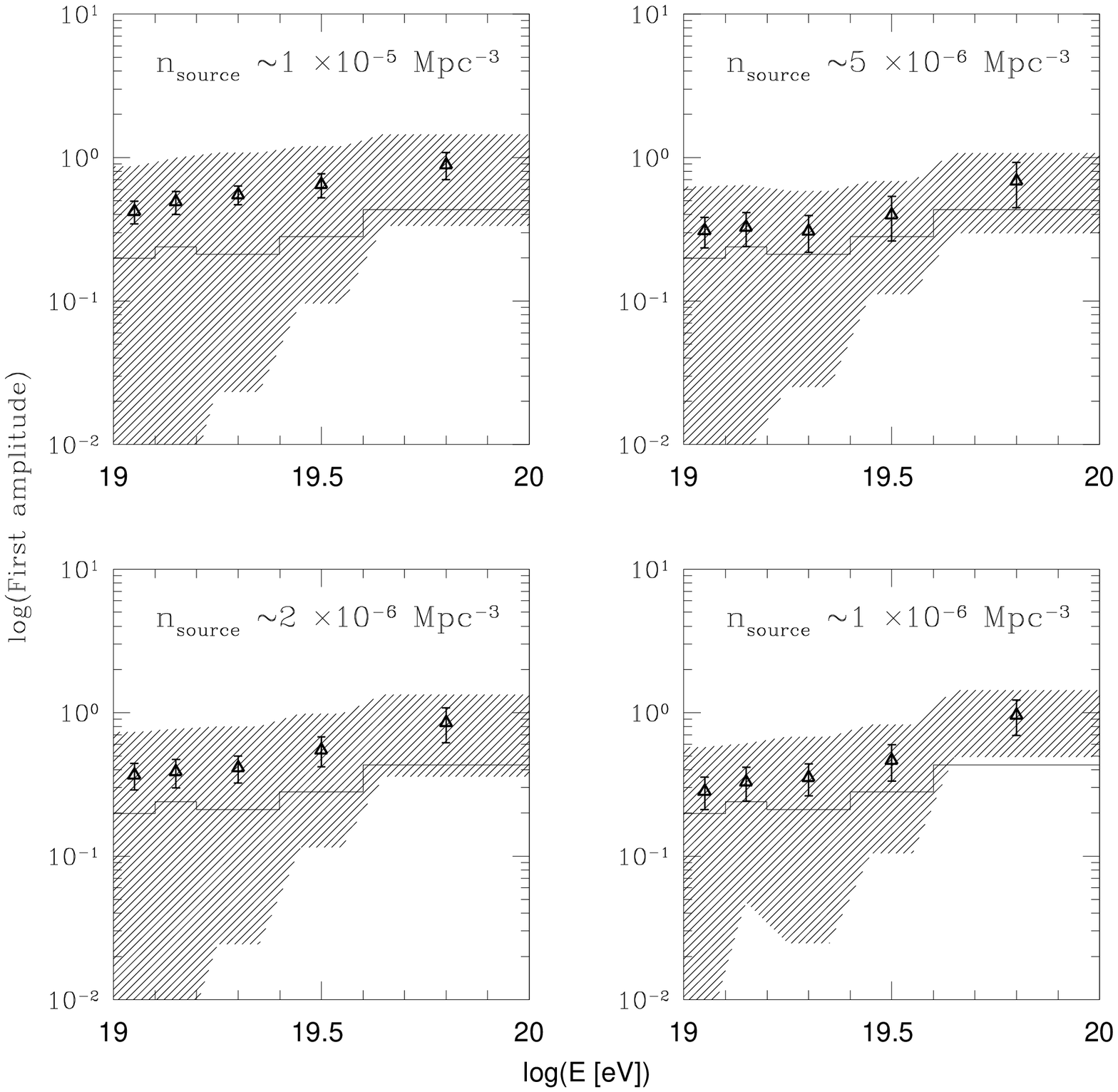} 
\caption{The first amplitudes predicted by our source models, as a function of the cosmic ray energies. The shaded regions represent 1 $\sigma$ total statistical error due to the event selection and the source selection. The errorbars represent the statistical fluctuations only due to the event selection. The number of simulated events is set to be equal to that observed by the AGASA. The upper left is the first amplitude calculated by a source model whose number density of source $n_{\mathrm{source}}$ is $\sim 1 \times 10^{-5}$ Mpc$^{-3}$. The upper right, the lower left and the lower right are the first amplitudes for $n_{\mathrm{source}} \sim 5 \times 10^{-6}, 2 \times 10^{-6}, 1 \times 10^{-6}$ Mpc$^{-3}$ respectively.}
\label{first}
\end{center}
\end{figure*}

\begin{figure*}
\begin{center}
\epsscale{1.2}
\plotone{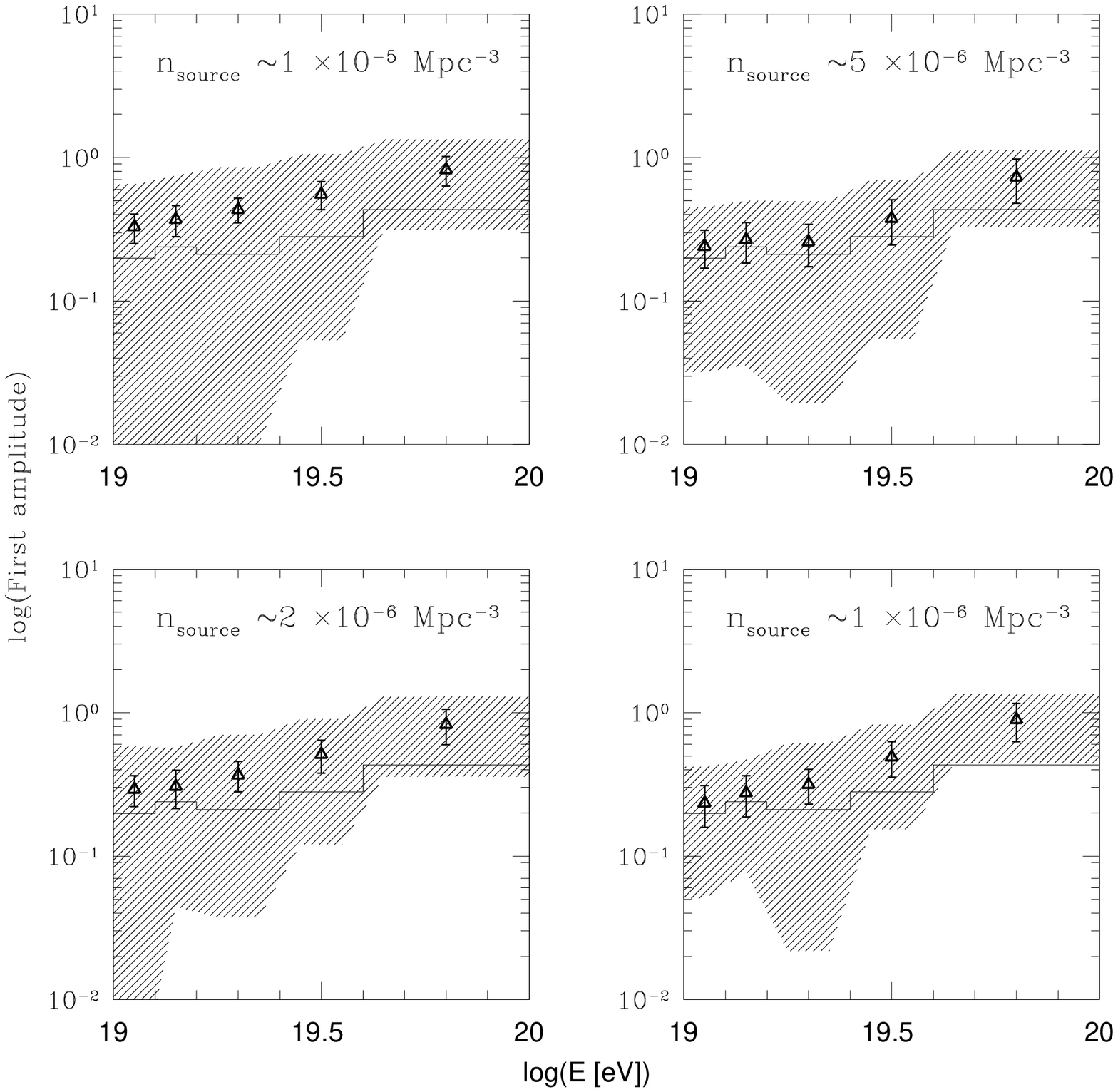} 
\caption{The same as Fig. \ref{first}, but for amplitudes of the second harmonics.}
\label{second}
\end{center}
\end{figure*}

Figure \ref{first} and figure \ref{second} show simulated harmonic amplitudes. The number of simulated events is set to be 775 in the energy above $10^{19}$ eV and their arrival direction is restricted in the range of $-10^{\circ} \leq \delta \leq 80^{\circ}$ in order to compare our results with those of the AGASA. Note that $\delta$ is the declination. The shaded regions represent 1 $\sigma$ total statistical error, which is caused by the two components of statistical error which occur from the finite number of simulated events and the random source selection from our IRAS sample. In order to see magnitudes of each error, we also draw errorbars, which represent the only statistical fluctuation due to the finite number of the simulated events. The event selection and the random source selection are performed 100 times and 40 times respectively. The regions below the solid lines are expected for the statistical fluctuation of isotropic source distribution with the chance probability larger than 10\%. For all source number density, both first and second amplitudes show that our source models predict sufficient isotropy of UHECR arrival distribution obtained by the AGASA.

Next, we investigate what number density of the sources reproduce the two point correlations obtained by the AGASA best. In order to evaluate it, we introduce $\chi_{\theta_{\mathrm{max}}}$ for a source distribution as 
\begin{equation}
\chi_{\theta_{\mathrm{max}}} = \frac{1}{\theta_{\mathrm{max}}} 
\sqrt{ \sum_{\theta = 0}^{\theta_{\mathrm{\mathrm{max}}}} 
\frac{\left\{ N(\theta) - N_{\mathrm{obs}}(\theta) \right\}^2}
{{\sigma(\theta)}^2}}, 
\label{511}
\end{equation}
where $N(\theta)$ is the two point correlation function calculated from simulated arrival distribution within $-10^{\circ} \leq \delta \leq 80^{\circ}$ and $N_{\mathrm{obs}}(\theta)$ is that obtained from the AGASA data at angle $\theta$. $\sigma(\theta)$ is total statistical error of $N(\theta)$ due to the finite number of simulated events. The random event selection are performed 100 times. This $\chi_{\theta_{\mathrm{max}}}$ represents goodness of fitting between the simulated two point correlation and the observed one. In this study, we take $\theta_{\mathrm{max}}$ to be 10$^{\circ}$.

\vspace{0.5cm}
\centerline{{\vbox{\epsfxsize=8.0cm\epsfbox{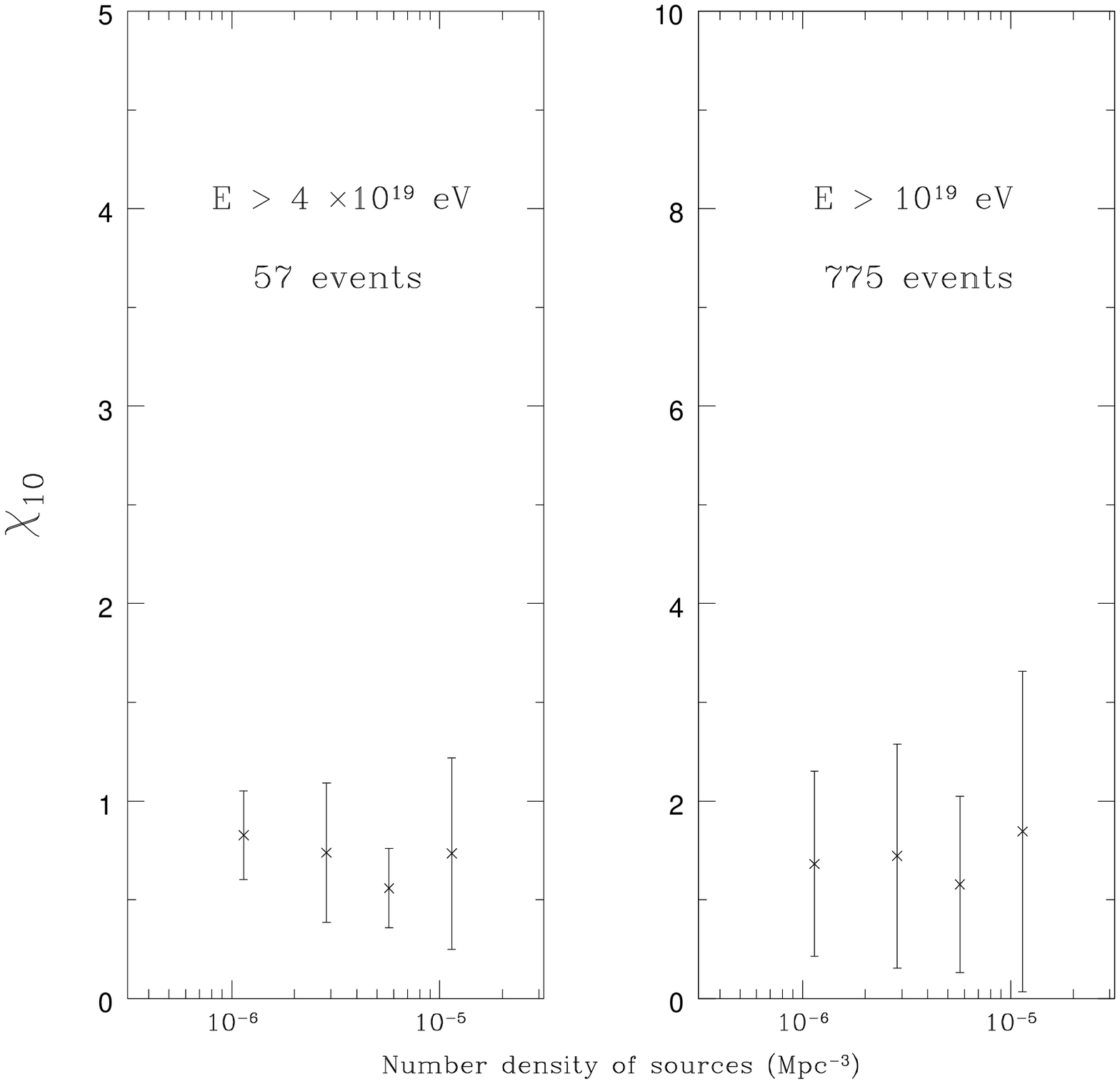}}}}
\figcaption{$\chi_{10}$ as a function of number density of source $n_{\mathrm{source}}$). The errorbars represent the statistical fluctuations due to the source selection from our galaxy sample. The left and right panel are calculated in the energy range of $E > 4 \times 10^{19}$ eV and $E > 10^{19}$ eV respectively.\label{chifit}}
\vspace{0.5cm}

Figure \ref{chifit} shows $\chi_{10}$ as a function of number density of the sources ($n_{\mathrm{source}}$). The errorbars represent the statistical fluctuations due to the random source selection from our galaxy sample. The random source selections are performed 40 times. The two point correlation functions of the left and right panel are calculated in the energy range of $E > 4 \times 10^{19}$ eV and $E > 10^{19}$ eV, respectively. However, we cannot know the normalization of two point correlation function obtained from the AGASA for $E > 10^{19}$ eV. Thus, we fit the two point correlation function obtained by the AGASA data to that calculated from our simulation at $\theta_{\mathrm{max}}$. 

Source models with larger source number density have strong peak at a small angle scale on the two point correlation since there are some sources near to the earth. On the other hand, in the case of smaller source number density, a small number of sources near to the earth contribute the arrival distribution, especially in the highest energy case (left panel), since radial distances between any two sources from the earth are more distant. Thus the peak of the two point correlation function also becomes more strong. Therefore $\chi_{10}$s should have a minimum as a function of source number density. As is seen from the figure, $n_{\mathrm{source}} \sim 5 \times 10^{-6}$ Mpc$^{-3}$ reproduces the two point correlation function obtained by the AGASA best.

Therefore, $n_{\mathrm{source}} \sim 5 \times 10^{-6}$ Mpc$^{-3}$ is the most appropriate number density of UHECR sources, since this source model also reproduces the harmonic amplitude obtained by the AGASA well. Note that number density of the sources have some uncertainty since the error bars in both panels are large.

Figure \ref{energy_spectrum} is the energy spectra at the earth predicted by our source models. These spectra are averaged ones among 40 source distributions on each source number density. Solid line represents energy spectrum obtained for $n_{\mathrm{source}} \sim 5 \times 10^{-6}$ Mpc$^{-3}$. We also show the observed cosmic-ray spectrum by the AGASA \citep*{hayashida00}. These simulated energy spectra have cutoffs around $E\sim 10^{19.6-8}$ eV except $n_{\mathrm{source}} \sim 1 \times 10^{-5}$ Mpc$^{-3}$, which are the GZK cutoff. These spectra can reproduce the AGASA data below $10^{20}$ eV. Note that the spectrum with $n_{\mathrm{source}} \sim 1 \times 10^{-5}$ Mpc$^{-3}$ has little spectral cutoff since there are many distributions which contain sources in the GZK sphere due to large source number density.  The source model ($n_{\mathrm{source}} \sim 5 \times 10^{-6}$ Mpc$^{-3}$) also reproduces the observed energy spectrum only below $10^{20}$eV. We concluded in \cite*{yoshiguchi03a} that a large fraction of cosmic rays above $10^{20}$eV observed by the AGASA might originate in the top-down scenarios. Thus we consider UHECRs with only $E < 10^{20}$eV in what follows. 

\vspace{0.5cm}
\centerline{{\vbox{\epsfxsize=8.0cm\epsfbox{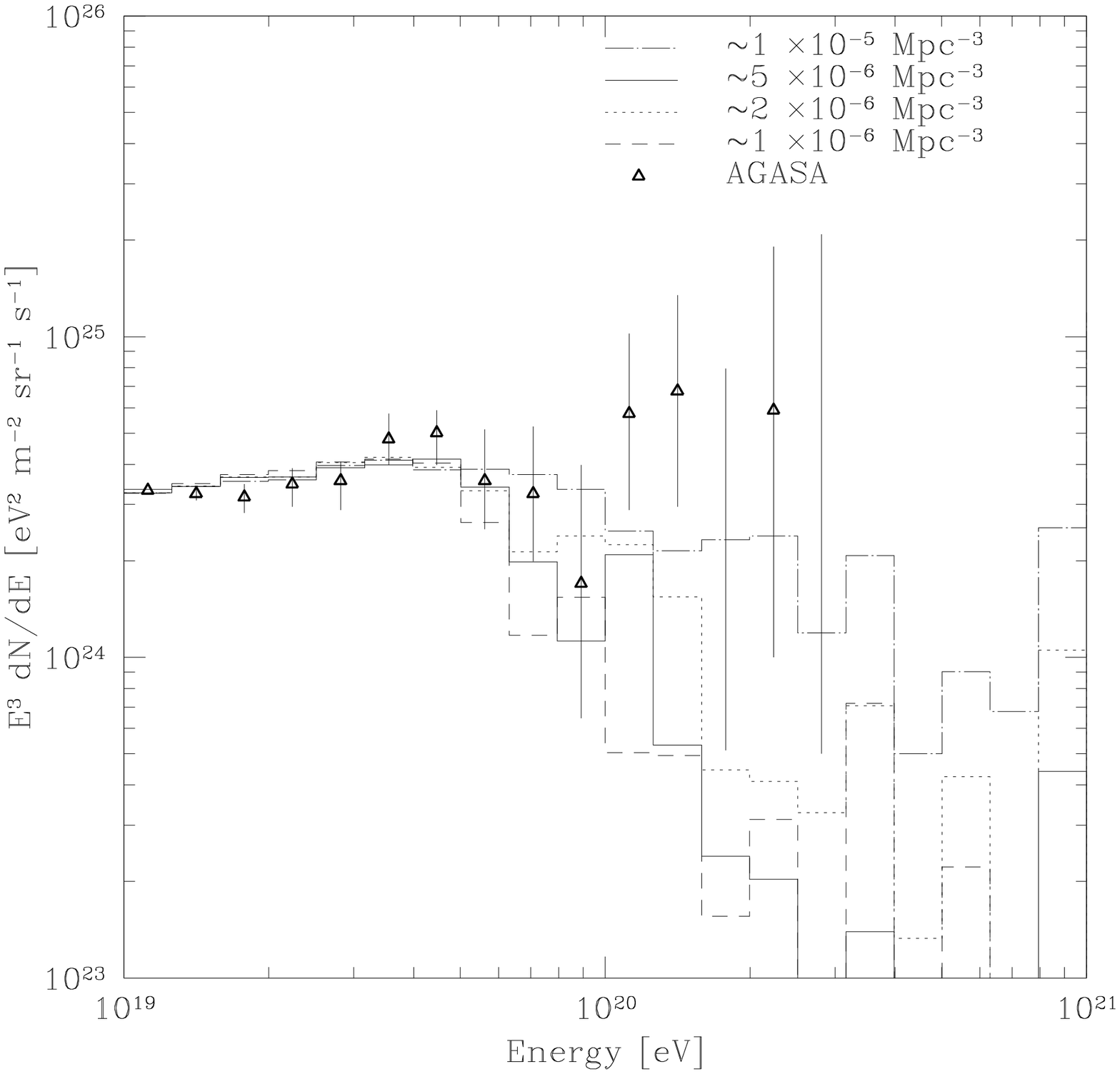}}}}
\figcaption{Energy spectra at the earth predicted by our source models. These spectra are averaged ones 40 times of the source selection on each source number density. Solid line represents energy spectrum obtained by a source model whose number density is $\sim 5 \times 10^{-6}$ Mpc$^{-3}$. We also show the observed cosmic-ray spectrum by the AGASA.\label{energy_spectrum}}
\vspace{0.5cm}

%%%%%%%%%%%%%%%%%%%%%%%%%%%%%%%%%%%%%%%%%
\subsection{Arrival Distribution of UHECRs above $10^{19}$eV} \label{arrival}
%%%%%%%%%%%%%%%%%%%%%%%%%%%%%%%%%%%%%%%%%

In this subsection, we demonstrate a skymap of the arrival distribution of UHECRs in the case of $n_{\mathrm{source}} \sim 5 \times 10^{-6}$ Mpc$^{-3}$. We construct their arrival distribution using our method explained in the section \ref{calc_arrival}. 

\begin{figure*}
\begin{center}
\epsscale{1.8}
\plotone{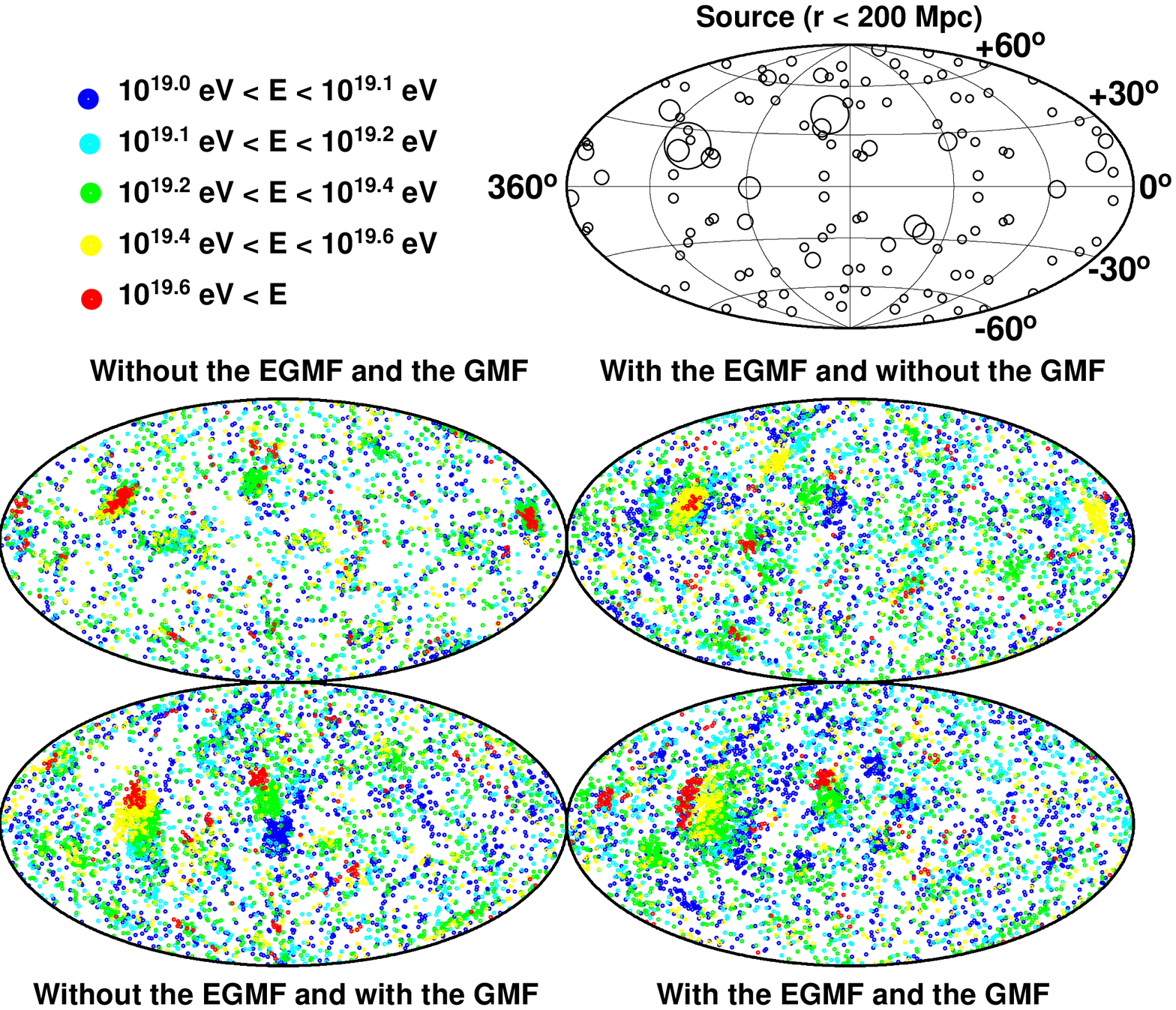} 
\caption{Skymaps of arrival distribution of UHE protons with $E > 10^{19}$ eV at the earth, which is expected for the source model of uppermost panel in the galactic coordinate. We show only the sources within 200 Mpc from us for clarity as circles of radius inversely proportional to their distances. The events are shown by color points according to their energies. The event number is 5000, which is the expected number of events observed by the Pierre Auger observatory \citep*{capelle98} for a few years. The upper left, upper right, lower right and lower left panel is calculated without any magnetic fields, with only the EGMF, with only the GMF and with both magnetic fields respectively.}
\label{arrival_map}
\end{center}
\end{figure*}

Figure \ref{arrival_map} shows one of results of the event generation above $10^{19}$eV calculated from a specific source model with $n_{\mathrm{source}} \sim 5 \times 10^{-6}$ Mpc$^{-3}$. The points represent each event and the events are colored according to their energies. The number of events is 5000, which is the expected number of events observed by the Pierre Auger observatory for a few years \citep*{capelle98}. The skymap generated with both the EGMF and the GMF is in the lower right panel and that without any magnetic fields, that with only the EGMF and that with only the GMF is in the upper left, the upper right and the lower left panel respectively. 

This specific source model has three strong sources (see upper right). One is $(l,b) \sim (199^{\circ}, 34^{\circ})$, another is $(l,b) \sim (287^{\circ}, 19^{\circ})$ and the other is $(l,b) \sim (25^{\circ}, 11^{\circ})$. Each distance from us is about 77 Mpc, 65 Mpc and 70 Mpc respectively. In the absence of any magnetic fields (upper left panel), there are the strong clusterings of events at the directions of these three sources. When the effects of the EGMF are included (upper right), we find the diffusion of the clustered events. In the lower left panel, the clustered events are arranged in the order of their energies, reflecting the directions of the GMF. This was pointed out by \cite*{stanev02} and \cite{yoshiguchi03d}. Note that we cannot find the clustered events at direction of one of the strong sources $(l,b)=(25^{\circ},11^{\circ})$. This is because UHE protons injected at this source cannot reach the earth due to the GMF. In the lower right panel, we also find the arrangements at the same points of the lower left panel. In addition, the EGMF diffuses these clustered events as we see in the two upper panels.

In order to see these features quantitatively, we compare the statistical quantities calculated with the EGMF to those calculated without the EGMF in the presence of the GMF in the next subsection.

%%%%%%%%%%%%%%%%%%%%%%%%%%%%%%%%%%%%%%%%%
\subsection{Statistics on the UHECR Arrival Distribution} 
\label{res_statistics}
%%%%%%%%%%%%%%%%%%%%%%%%%%%%%%%%%%%%%%%%%

In this subsection, we compare the statistical quantities on the arrival distribution calculated with the EGMF to those without the EGMF. We take $n_{\mathrm{source}} \sim 5 \times 10^{-6}$ Mpc$^{-3}$.

Figure \ref{two57} and figure \ref{two775} show the two point correlation functions simulated by our source model in the energy range of $E > 4 \times 10^{19}$eV and $E > 10^{19}$eV respectively. In each figure, the left panel shows the two point correlation function calculated with the EGMF and the right panel shows that without the EGMF. Note that the GMF is taken into account in the figures. We calculate the two point correlation function for the simulated events within only $-10^{\circ} \leq \delta \leq 80^{\circ}$ in order to compare our results with the AGASA data. The shaded regions represent 1 $\sigma$ total statistical error, which is caused by the finite number of simulated events and the random source selections from our IRAS sample. We also draw errorbars, which represent the statistical fluctuations due to the finite number of the simulated events, which is set to be equal to that observed by the AGASA (49 events for $E > 4 \times 10^{19}$ eV and 775 events for $E > 10^{19}$ eV). The event selection and the source selection are performed 100 times and 40 times respectively. The histograms represent the AGASA data \citep*{takeda01}. For the data of $E > 10^{19}$eV, we normalize the two point correlation function as the correlation function obtained by the AGASA fits the calculated one at 30$^{\circ}$, since we cannot know the normalization of the AGASA data with this energy. 

In both figure \ref{two57} and \ref{two775}, it is visible that a peak at small angle is much stronger than that of the AGASA though the AGASA data were covered in the shaded regions, which are mainly caused by the source selection. In our previous work \citep*{yoshiguchi03d} in which the EGMF was neglected, we also faced this situation and pointed out possible explanations, one of which was effects of the EGMF. 

In figure \ref{two775}, we find that a peak of two point correlation function calculated with the EGMF at small angle scale is weaker than that without the EGMF. Thus the consistency with the AGASA data becomes better due to the EGMF. On the other hand, in figure \ref{two57}, a obvious difference between the two panels on the peak at small angle is not found. This is because UHECRs above $4 \times 10^{19}$ eV are less deflected by the EGMF and hardly dispersed. 

\vspace{0.5cm}
\centerline{{\vbox{\epsfxsize=8.0cm\epsfbox{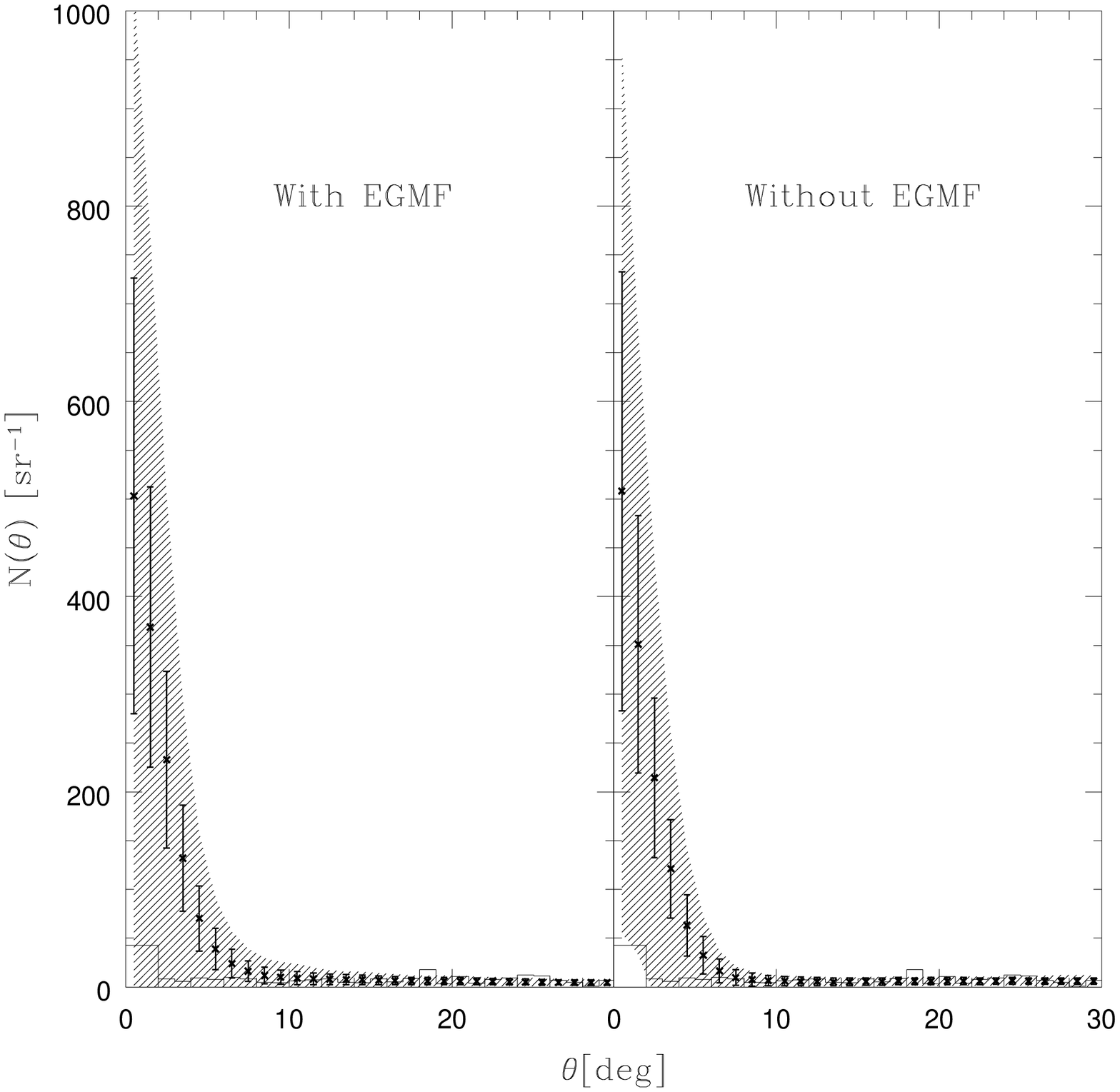}}}}
\figcaption{Two point correlation functions calculated from our source model with $n_s \sim 5 \times 10^{-6} \mathrm{Mpc}^{-3}$ for $E > 4 \times 10^{19}$ eV (49 events). The left panel shows the two point correlation calculated with the EGMF and the right one shows that calculated without the EGMF. The shaded regions represent 1 $\sigma$ total error to consist of statistical error due to the finite number of the simulated events, which is set to be equal to that observed by the AGASA within $-10^{\circ} < \delta < 80^{\circ}$, and one due to the source selection. The errorbars represent the statistical fluctuation due to the finite number of the simulated events in order to see this error contribute to the total error. The histograms represent the AGASA data.\label{two57}}
\vspace{0.5cm}

\vspace{0.5cm}
\centerline{{\vbox{\epsfxsize=8.0cm\epsfbox{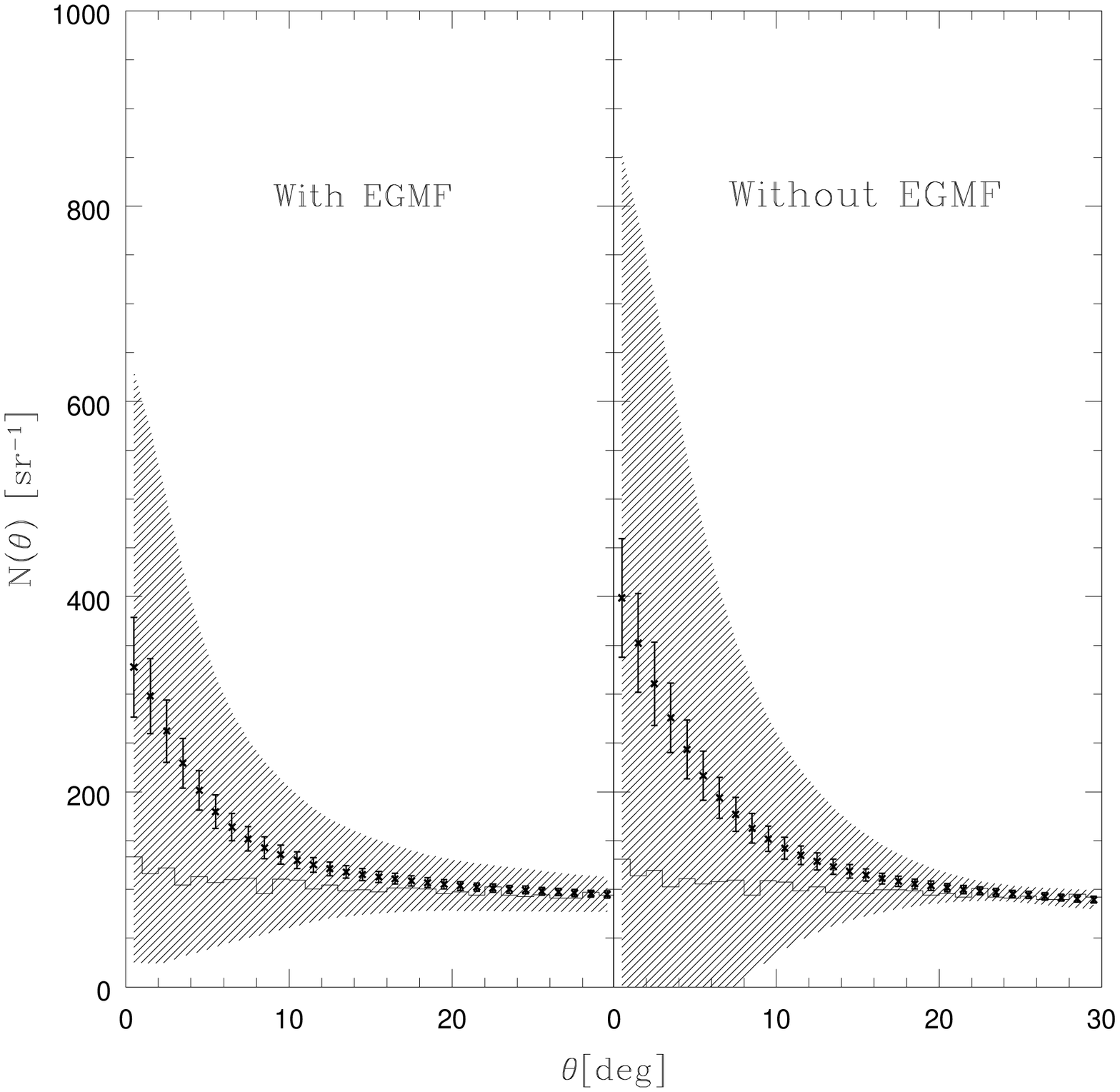}}}}
\figcaption{Same as figure \ref{two57}. But these two point correlation functions are calculated for $E > 10^{19}$ eV (775 events within $-10^{\circ} < \delta < 80^{\circ}$ ). We fit the AGASA data to that calculated from our source model at $\theta_{\mathrm{max}} = 30^{\circ}$ since we cannot know the normalization of the AGASA data with this energy range.\label{two775}}
\vspace{0.5cm}

As mentioned above, the calculated two point correlation functions have large errors since some source distributions out of 40 contain very near source in our source model. Such source distributions do not reproduce the large-scale isotropy observed by the AGASA. We check that 20 source distributions out of 40 predict the sufficient large-scale isotropy obtained by the AGASA within 1 $\sigma$ statistical error due to the finite number of the simulated events. From these 20 source distributions, we calculate the two point correlation function in the presence of the EGMF again. The results are shown in figure \ref{two_well}.

\vspace{0.5cm}
\centerline{{\vbox{\epsfxsize=8.0cm\epsfbox{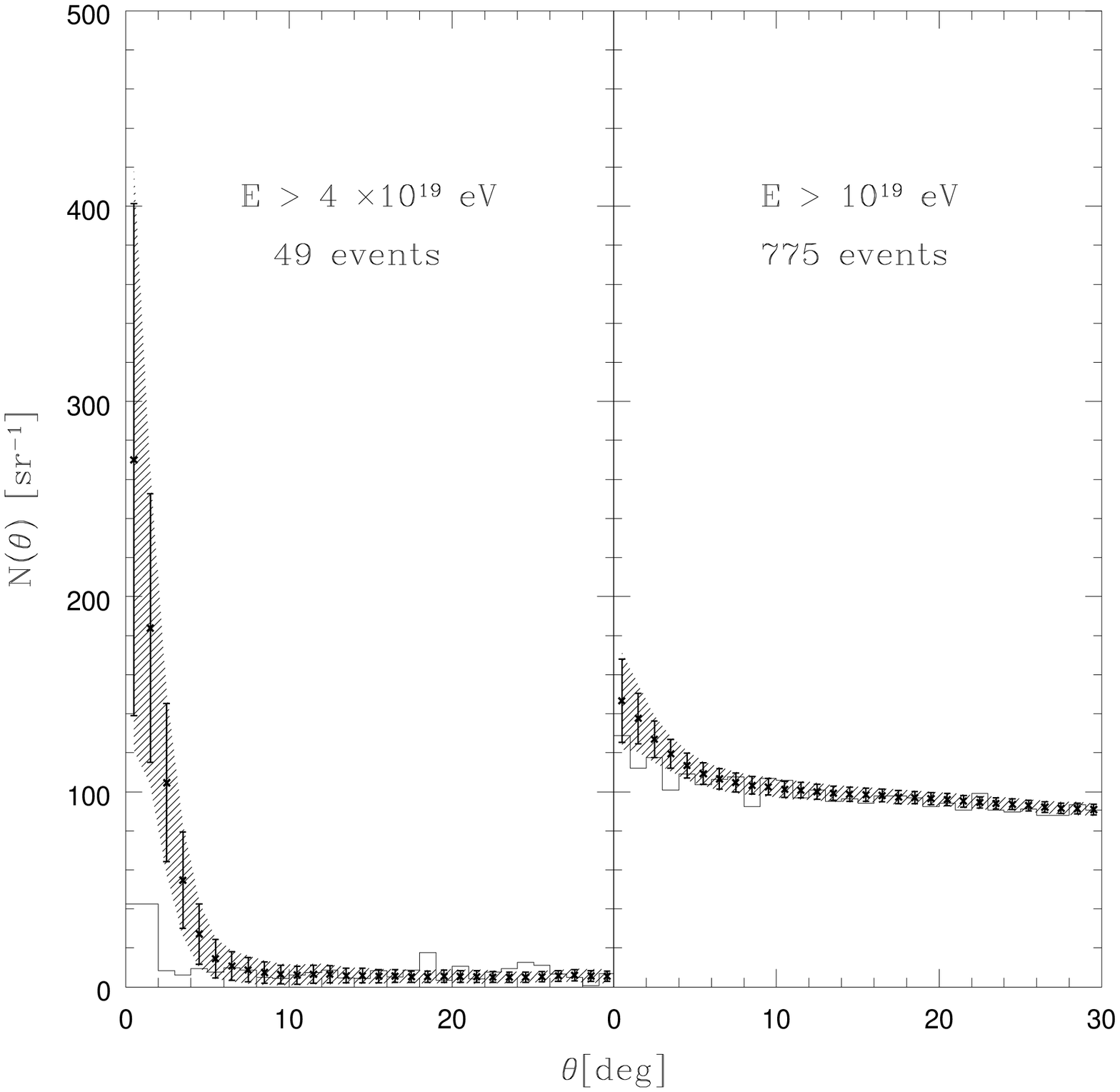}}}}
\figcaption{Two point correlation functions calculated with the EGMF from 20 source distribution which satisfy the large-scale isotropy observed by the AGASA sufficiently, for $E > 4 \times 10^{19}$ eV (49 events, left) and $E > 10^{19}$ eV (775 events, right). These functions are calculated within $-10^{\circ} < \delta < 80^{\circ}$. The shaded regions and the errorbars have the same mean as figure \ref{two57}, \ref{two775}. The histograms represent the AGASA data. The normalization of the AGASA data with $E > 10^{19}$ eV performs the same as figure \ref{two775}.\label{two_well}}
\vspace{0.5cm}

In figure \ref{two_well}, we find that the consistency with the AGASA data becomes better than the two point correlation functions in figure \ref{two57} and \ref{two775}. We also see that the errors due to the source selection become small. However, we should note that the peak at small angle scale is still relatively strong,  compared with the AGASA though our previous result \citep*{yoshiguchi03d} is improved by the effect of the EGMF. We assume effects of the random component of the GMF, which is neglected in this work, as one of possible explanations for this fact. This issue is left for future investigations.

We also investigate the harmonic amplitudes in the same way. But there
is little difference dependent on the EGMF. 

%%%%%%%%%%%%%%%%%%%%%%%%%%%%%%%%%%%%%%%%%%%%%%%%%
\section{SUMMARY AND DISCUSSION} \label{summary}
%%%%%%%%%%%%%%%%%%%%%%%%%%%%%%%%%%%%%%%%%%%%%%%%%

We presented numerical simulations on propagation of UHECRs above $10^{19}$ eV in a structured EGMF and GMF. We used the IRAS PSCz catalogue in order to construct a structured EGMF model which reflects the local structures actually observed, and source models of UHECRs. We calculated an inverse process of their propagation taking the energy loss processes into account in the EGMF. We injected UHECRs from the earth isotropically whose charges are taken as -1 and recorded these trajectories. They could be regarded as trajectories of UHE protons from the extragalactic space. We then select some of their trajectories according to given source distributions. The simulated arrival distribution was able to be obtained by mapping the velocity directions of the selected trajectories at the earth. The use of this method enabled us to calculate only trajectories of UHECRs reaching the earth and saved the CPU time effectively. The validity of this method was supported by the Liouville's theorem. 

We calculated the harmonic amplitudes and the two point correlation functions of arrival distribution of UHECRs above $10^{19}$eV, using our source models and examined what number density of the sources reproduces the large-scale isotropy and the small-scale anisotropy obtained by the AGASA best. As a result, we found that $\sim 5 \times 10^{-6}$ Mpc$^{-3}$ was the most appropriate number density of source of UHECRs. Number density of source is a constraint on source candidate of UHECRs.

We also demonstrated skymaps of the arrival distribution of UHECRs above $10^{19}$ eV, using our source model for $n_{\mathrm{source}} \sim 5 \times 10^{-6}$ Mpc$^{-3}$  with the event number expected by future experiments and examined how the EGMF affects the arrival distribution of UHECRs. The main result was diffusion of clustering events which is obtained by calculations neglecting the EGMF. 

In order to see the effect of the EGMF quantitatively, we compared the two point correlation function calculated with our structured EGMF model to that without the EGMF. We found that the EGMF weakened the small-scale anisotropy and improved a prediction in \cite*{yoshiguchi03d}, which had been calculated with only the GMF.

However, we found the calculated two point correlation functions had large errorbars since source distributions, which contained sources very near to us, existed. Such source distributions do not reproduce the large-scale isotropy observed by the AGASA. Thus we calculated the two point correlation functions from source distributions which predicted the large-scale isotropy obtained by the AGASA within 1 $\sigma$ statistical error. These simulated two point correlation function reproduced that of the AGASA better.  It is possible that these functions at small angle scale can be closer to the observational data due to the random component of the GMF. This issue is left for future studies. 

New large aperture detectors are under development, such as the Pierre Auger observatory \citep*{capelle98}, the EUSO \citep*{euso92} and the Telescope Array. These projects are expected to increase observed events of UHECRs per year drastically. We can obtain more strong constraints on our source model, other than number density of sources, using other statistical quantities when the detailed data of large events of UHECRs are published. This is one of plans of future studies.

%%%%%%%%%%%%%%%%%%%%%%%%%%%%%%%%%%%%%%%%%%%%%%%%%%%%%%%%%%%%%%%
%%%%%%%%%%%%%%%%%%%%%%%%%%%%%%%%%%%%%%%%%%%%%%%%%%%%%%%%%%%%%%%
\acknowledgments
%%%%%%%%%%%%%%%%%%%%%%%%%%%%%%%%%%%%%%%%%%%%%%%%%%%%%%%%%%%%%%%
%%%%%%%%%%%%%%%%%%%%%%%%%%%%%%%%%%%%%%%%%%%%%%%%%%%%%%%%%%%%%%%

The work of H.Y. is supported by Grants-in-Aid for JSPS Fellows. 
The work of K.S. is supported by Grants-in-Aid for Scientific Research 
provided by the Ministry of Education, Science and 
Culture of Japan through Research Grant No.S14102004 and No.14079202.

%Grant number is unknown.

%%%%%%%%%%%%%%%%%%%%%%%%%%%%%%%%%%%%%%%%%%%%%%%%%%%%%%%%%%%%%%%%%%%%%%%
%\clearpage
%%%%%%%%%%%%%%%%%%%%%%%%%%%%%%%%%%%%%%%%%%%%%%%%%%%%%%%%%%%%%%%%%%%%%%%

%%%%%%%%%%%%%%%%%%%%%%%%%%%%%%%%%%%%%%%%%%%%%%%%%%%%%%%%%%%%%%%%%%%%%%%

\end{document}